# MontiCore: Modular Development of Textual Domain Specific Languages

Holger Krahn, Bernhard Rumpe, and Steven Völkel

Institute for Software Systems Engineering
Technische Universität Braunschweig, Braunschweig, Germany
`http://www.sse-tubs.de`

**Abstract.** Reuse is a key technique for a more efficient development and ensures the quality of the results. In object technology explicit encapsulation, interfaces, and inheritance are well-known principles for independent development that enable combination and reuse of developed artifacts. In this paper we apply modularity concepts for domain specific languages (DSLs) and discuss how they help to design new languages by extending existing ones and composing fragments to new DSLs. We use an extended grammar format with appropriate tool support that avoids redefinition of existing functionalities by introducing language inheritance and embedding as first class artifacts in a DSL definition. Language embedding and inheritance is not only assisted by the parser, but also by the editor, and algorithms based on tree traversal like context checkers, pretty printers, and code generators. We demonstrate that compositional engineering of new languages becomes a useful concept when starting to define project-individual DSLs using appropriate tool support.

## 1 Introduction

Reuse of developed artifacts is a key technique for more efficient development and high quality of the results. This is especially the case for object oriented programming: existing well-tested code is packed into libraries in order to reuse the developed components in new projects. However, these principles are not consequently applied when designing languages in general and domain specific languages (DSLs) in particular. Those languages are often built from scratch without explicit and systematic reuse of existing languages or fragments beyond some knowledge on language design in the heads of designers.

The idea of a DSLs is to assist the tasks of developing software through efficient development of models. Based on DSL models, property analysis, metrics, and smells as well as code generation help the developers to become more efficient and to deliver higher quality systems. Today, when applying DSL-based software development it takes too long to create all these tools, because they are implemented from scratch too often. Thus, we explain an infrastructure that allows reuse, extension, and composition of existing languages and their tools.

Our experience in some projects where we helped developers designing a language specific for their domain shows that the creation of high-quality languages is a labor intensive task. Hence, once a language is developed, the reuse in other contexts is highly desirable. This idea was discussed in [Spi01] in form of design patterns for the development of DSLs. Different strategies such as language extension, language specialization, or piggypack (language combination) are introduced and preferred to a standalone realization. However, most of today's DSL frameworks do not support these patterns and so methodological assistance in this respect is poor.

In this paper we present work on appropriate tooling based on the MontiCore framework [GKR+06, KRV07b, Mon], which allows defining a DSL by an integrated specification of abstract and concrete syntax in a concise grammar-based format. It supports rule inheritance, introduction of interfaces, and associations directly in the grammar which results in an abstract syntax that outbalances the common tree structure and is comparable to current metamodeling standards. Because modularity and existing composition techniques are core requirements for reuse, we explore two mechanisms for modularizing grammars. We apply *language inheritance* that can be used in order to extend existing languages by redefining productions or adding alternatives in conjunction with the introduction of interfaces in a language definition. In addition, *language embedding* can be used to define explicit nonterminals in a grammar which can be filled by a fragment of another language (e.g., expressions or statements). Most importantly, this can be done at configuration time, and thus allows a modular independent development and compilation of tools that deal with language fragments. Because both, guest and host languages have separate lexers and parsers, they can be developed independently and do not interfere with each other. This technical separation allows a component based composition of grammars and their tools. Thus we are able to set up libraries of quality assured languages which can be reused in other contexts.

Both modularity mechanisms are implemented in MontiCore. In this paper we explain how these mechanisms are integrated in generation of language recognition artifacts such as parsers and lexers as well as the abstract syntax in the form of a strongly typed abstract syntax tree (AST). Furthermore, language specific editors with several comfort functionalities such as syntax highlighting or outlines can be generated as Eclipse plugins in a modular fashion to support an efficient use of the language under design [KRV07a].

The rest of this paper is structured as follows: Section 2 explains existing approaches from compiler design and metamodeling to design language based tools in a modular fashion. Section 3 describes the basic syntax of the MontiCore grammar format. In Section 4 two different concepts for defining abstract and concrete syntax in a modular way are explained. Section 5 describes how this modular language definition permits other functionalities like tree traversal and editor generation to be specified in a modular fashion. Section 6 concludes the paper.

## 2 Related Work

Modularity and composing complex systems from modules [Par72] is an important concept in computer science because it enables multiple developers to work concurrently on the same project. It allows them to understand the system part under design without requiring them to understand the whole system.

In [Spi01] different design patterns are introduced. Especially *language extension* and *piggyback* describe two modularization mechanisms. *Language extension* describes the extensions of a host language (often an existing GPL) by elements that are domain specific. This approach for specifying domain specific languages (DSL) is often named *embedded* DSLs. *Piggyback* describes the extension of a DSL by extracts of other languages, for example, GPL statements in a DSL for describing context free grammars. This combination is often used as an input language for parser generators.

In the context of grammar based software (for short: grammarware [KLV05]) the modular development of parsers is an important goal. The class of context free grammars is closed under composition in the sense that the composition of two context free languages is again a context free language. The main problem is that this property does not hold true for the subsets that are usually used for parsing like LL(k) or LR(k). To solve this problem, more sophisticated parsing algorithms like Generalized-LR [Tom85] and Early parsers [Ear70] have been created. Packrat parsing [For02, Gri06] uses parsing expression grammars which are closed under composition, intersection, and complement.

A particularly difficult problem is the composition on the lexical level. In [BV07] possible solutions are discussed. The concrete solution proposed for this problem is scannerless parsing where no separate lexer exists, which might in turn impose runtime difficulties depending on the composed languages. In the following we show that the control of the lexer state from the parser, that was not favored by the authors of [BV07] for technical reasons, can be seamlessly integrated in a DSL framework.

The focus of language libraries discussed in [BV07] is to realize embedded DSLs where the guest language is assimilated to the host language in order to design the extension in a modular way. MetaBorg [BV04, BdGV05] uses a GPL as a host language. The DSL is assimilated to the GPL functionality by mapping the DSL code to library code that provides the desired functionality. In contrast, our approach does not specifically aim at embedded DSLs but at the combination of separately developed DSLs that are all mapped to a GPL by separate but cooperating code generations. For this kind of problem the assimilation phase does not apply because due to their restrictive expressiveness usually different DSLs cannot be assimilated to each other. Embedded DSLs can also be realized with Attribute Grammars that focus on how distinct attributes can be realized in a modular way (e.g., using Forwarding [WMBK02]).

Even though the above mentioned parsing algorithms can achieve compositionality, an intrinsic problem of language composition is to avoid ambiguities in the composed language. Although GLR and Early can, depending on the implementation, return a set of all ASTs instead of a single one, further development steps like semantic analysis and code generation usually require the choice of exactly one tree. Packrat parsing does return at most one tree, as ambiguities are avoided by design. The alternatives are prioritized and limited backtracking functionality is possible by using predicates. Therefore, the developer has to pay close attention to the order of alternatives when designing a language. In general for any given parsing algorithm, ambiguities must be avoided and therefore, good language composition shall help the user to detect ambiguities and help him to circumvent them.

In the field of attribute grammars, modularity is a highly researched topic. Multiple language inheritance [MŽLA99, MLAŽ99] helps to extend existing language and attribute definitions. Some approaches use generic patterns to create reusable grammar chucks (e.g., [Ada91]). A few compiler frameworks, e.g., JastAdd [EH07], focus on modular extensible language definitions and compiler construction.

The focus of compiler design and DSL frameworks is a bit different. Compiler frameworks and related tools usually aim at programming languages and their modular extensions. An important property is that two extensions of a base language can be developed independently of each other and be integrated seamlessly. In this way, embedded DSLs are realized. The frameworks target at a single form of code generation usually towards the host language or bytecode. On the contrary, DSL frameworks focus on the creation of modeling languages that are not necessary executable. The DSLs are used for a variety of purposes like product and test code generation, documentation, and model-to-model transformations.

DSL frameworks (e.g., [LMB+01, Metb]) often rely on a graphical concrete syntax and do not support the user with a modular language definition. Other tools like [Meta] allow defining textual DSLs but do not provide modularity concepts either. DSL frameworks like OpenArchitectureWare [Ope] based on EMOF or Ecore and Moflon [AKRS06] based on CMOF use package imports and merges to allow a compositional definition of the abstract syntax of a language. The definition of the concrete syntax using tools like TCS [JBK06] or xText [Ope] are usually not compositional.

## 3 Language Definition using MontiCore

MontiCore uses an enriched context-free grammar as input format which is similar to the input format of Antlr [PQ95] that is also used for parser generation. Figure 1 contains an illustrative example of a DSL describing a bookstore.

The grammar body consists of regular lexer and context-free parser rules. Lexer rules are marked by the keyword `ident` and defined by regular expressions. To simplify the development of new languages, we use predefined rules for `IDENT` and `STRING` to recognize names and strings. A language may declare its own lexer rules as shown in line 7. In addition, a mapping to a Java type can be defined. In our case, `ID` is mapped to the built-in type `int`. More complex mappings can be specified using Java directly [KRV07b].

──────────────── MontiCore-Grammar ────────────────

```
1  package mc.examples.bookstore;
2
3  grammar Bookstore {
4
5    // Create a token "ID" which is reflected
6    // as int in the abstract syntax
7    ident ID ('0'..'9')+ : int;
8
9    Bookstore = "bookstore" name:IDENT "{" ( Book | Journal )* "}" ;
10
11   Book = "book" id:ID title:STRING "by"
12            authors:Person ("," authors:Person)* ";" ;
13
14   Journal = "journal" id:ID title:STRING ";" ;
15
16   Person = forename:IDENT lastname:IDENT ;
17
18 }
```

**Fig. 1.** Bookstore example

Parser rules have a name and a right-hand-side (RHS) which describes the syntactical structure recognized by the generated parser as well as the structure of the abstract syntax. A RHS consists of references to other lexer or parser rules, alternatives that are separated by "|", and blocks which are surrounded by brackets (e.g., line 9). Furthermore, optional elements can be expressed by a question mark, a Kleene star denotes unbounded repetitions, and a plus denotes a cardinality of at least one. In addition, we use a package mechanism similar to Java: grammars have a name (`Bookstore` in line 3) and an optional package (line 1) which determines the fully qualified name and the desired location in the file system.

MontiCore automatically derives an abstract syntax from this grammar format as follows. Each production forms a class having the same name as the production. Each reference to a lexer rule on the RHS forms an attribute of this class. References to parser rules are reflected as composition relationships with automatically determined cardinalities. In addition, the references can explicitly be named (e.g., line 12: `authors:Person`) in order to define the name of the attributes and compositions. Figure 2 shows the abstract syntax derived for our bookstore-example. Note that the abstract syntax is automatically mapped to Java classes in the same package as the grammar. Get- and set-methods for attributes as well as for compositions, tree traversal support, clone-methods, and equal-methods are automatically generated to simplify the use of the abstract syntax.

The abstract syntax is not limited to trees but additional associations between classes of the abstract syntax can be specified in the language definition. The linking of the objects is established after parsing the tree structure. The al-gorithms responsible for linking can either be specified in a declarative manner or can be hand-programmed depending on the suitability of standard mechanisms for references within a DSL. For deeper discussions on the grammar format as well as on the derivation of the abstract syntax and especially on the associations we refer to [KRV07b].

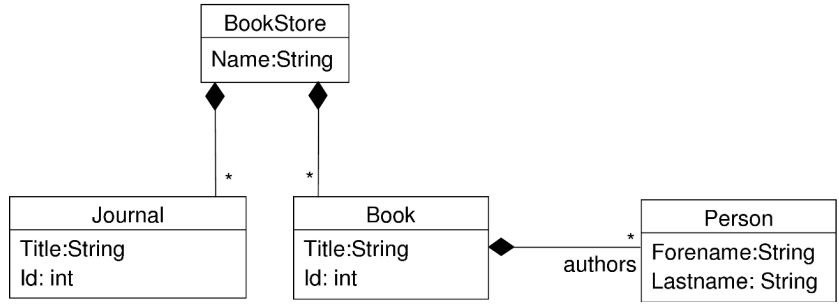

**Fig. 2.** Abstract syntax for the bookstore example defined in Figure 1

Typically counter-arguments exist against the decision of specifying abstract and concrete syntax in one format. We think that the most important pro argument is that no inconsistencies can occur between both artifacts which simplifies the development of a DSL. This cost reduction is of high importance as the initial investment for defining a DSLs has often to be proven worth by overall reduced project costs and quality improvements.

We are aware of tool infrastructure that can keep both artifacts consistent (like [KW96]) but think that the unified format also helps the user to keep both artifacts aligned with each other quite closely. The advantage is that the structure of the language (and based on that, its semantics) is aligned with the user perception of it (= the concrete syntax).

More technically, the most common counter-arguments are: First, there is a demand for multiple concrete syntaxes for a single abstract syntax. We doubt that this is the case for DSLs where the concrete syntax usually emerges from the domain. Often this argument targets at the reuse of development artifacts like code generation, but as model-to-model transformation facilities have emerged in the last years to a mature state, more explicit and flexible ways to reuse code generations for multiple languages than using a common concrete to abstract syntax mapper exist. Second, the abstract syntax is then not abstract enough, as syntactic sugar, associativity, and priorities of operators are still presented in the abstract syntax. This is not the case in MontiCore, as we extended the grammar format in such a way that multiple rules can refer to the same class in the abstract syntax. However, we will not discuss this feature in detail here.

MontiCore generates parsers that are based on Antlr 2.74. Therefore, the used parsing algorithm is LL(k) with semantic and syntactic predicates which is sufficient to parse a large variety of languages. We extended the parser generator by adding the generation of heterogeneously and strongly typed AST-classes including inheritance and associations. The main criteria for choosing Antlr in re-spect to other parser generators mentioned in Section 2 was the relatively concise EBNF based syntax that makes it attractive to inexperienced language engineers. The resulting parsers perform well and also due to the classical lexer/parser distinction standard algorithms for error messages and recovery strategies are well understood. The recursive descent structure of generated parsers makes it easy to debug the parsers and feasible for us to integrate the modularity concepts and the building of our abstract syntax as explained in this paper. In addition, every production can be used as a starting production which helps to reuse fragments in other composed languages.

## 4 Concepts of Modularity

In software engineering, the most commonly used modeling language is UML which is often combined with OCL for expressing constraints or actions. Taking this combination as an example for our language modularization mechanisms, we can make three main observations. First, a core which contains elements that are used in most of the UML sublanguages (e.g., classes or stereotypes) can be identified. It would be reasonable to extract these elements into a core language and to reuse it in other sublanguages like the UML suggests with its modular definition in different packages. Second, one could imagine using another constraint or action language than OCL for UML. Third, OCL could be combined with other languages when there is a demand for a constraint sublanguage. To summarize, a tight coupling between UML and OCL is not desirable when designing such a modeling language.

The reuse of parts of a language in a different context and a loosely coupled combination of languages are supported by MontiCore by its modularity concepts: *language inheritance* and *language embedding*. The former can be used to define a grammar by extending one or more supergrammars whereas *language embedding* permits to define explicit nonterminals in grammars which can be filled at configuration time by an appropriate embedded language. Both mechanisms are introduced in the following; we will discuss their effects especially on the abstract syntax and the advantages for defining new languages.

### 4.1 Language Inheritance

Language inheritance can be used when an existing DSL shall be extended by defining new productions or existing nonterminals shall be overridden without modification of the supergrammar. Therefore, the extending grammar defines only the differences between the existing language and the new one. We use the concept of multiple language inheritance as introduced in [MŽLA99] for attribute grammars. It can be seen as a method to achieve *Language extension* or *Language specialization* as discussed in [Spi01]. *Language extension* is typically the case when the subgrammar adds new alternatives to an existing rule by overriding it. A well-known example for language extension is LINQ [MBB06] where SQL-statements can be used as an expression inside an existing general purpose language. *Language specialization* occurs, e.g., by adding additional context constraints. It is often used to remove "unsafe" features of a language to gain safe sublanguages. In both cases, the definition of new languages is not desirable. Instead, a reuse should be preferred.

The MontiCore grammar format allows developers to define language inheritance by using the keyword `extends` followed by a list of fully qualified grammar names in the header of a grammar. From the concrete syntax point of view, the nonterminals and terminals of all supergrammars are visible in the current grammar and can therefore be reused. Furthermore, MontiCore enables to override existing productions by specifying a production with the same name. In contrast to [MŽLA99] we use an ordered inheritance approach where in the case of name collisions, i.e. two supergrammars use a common production name, the production from the first supergrammar is used. For example if both grammars `A` and `B` share a production name `X` and grammar `C` inherits from both `A` and `B`, the production from `A` is used. As an extension to the current implementation we plan to integrate a more sophisticated mechanism than the order of supergrammars to resolve conflicts. A simple example is shown in Figure 3 where the nonterminal `Journal` is redefined by adding editor information.

---
———————————————— **MontiCore-Grammar** ————————————————

```
1  package mc.examples.bookstore2;
2
3  grammar ExtendedBookstore extends mc.examples.bookstore.Bookstore {
4
5    Journal = "journal" id:ID title:STRING "editors"
6             editors:Person ("," editors:Person)* ";" ;
7
8  }
```
---

**Fig. 3.** Definition of Language Inheritance in MontiCore

The inheritance on the grammar basis leads to a modification of the abstract syntax of the language that is comparable to a package merge [OMG05]. For each overridden production a new class is created that inherits from all classes that are associated with the productions with the same name in the supergrammars. The key difference is that all unmerged classes remain unchanged and are not directly present in the merged package which is the package of the subgrammar. We found this approach appealing because algorithms such as transformations, code generators, or even symbol tables written for the original language still work for the extended languages (maybe with minor modifications for overridden productions).

The resulting abstract syntax for our running example is outlined in Figure 4. The new class `Journal` inherits from the version of its supergrammar. Please note that the new generated parser produces instances which conform to the new grammar, i.e., there will be only instances of `mc.examples.bookstore.Book`,

`mc.examples.bookstore2.Journal`, `mc.examples.bookstore.BookStore`, and `mc.examples.bookstore.Person`. Objects of `mc.examples.bookstore.Journal` will be not be created.

*Language inheritance* can lead to multiple inheritance of classes in the abstract syntax which is a problem in the current MontiCore implementation. The upcoming generation of AST-classes will map the classes of the abstract syntax to Java interfaces and create implementations like other metamodeling tools [BSM+03, AKRS06] to avoid such problems.

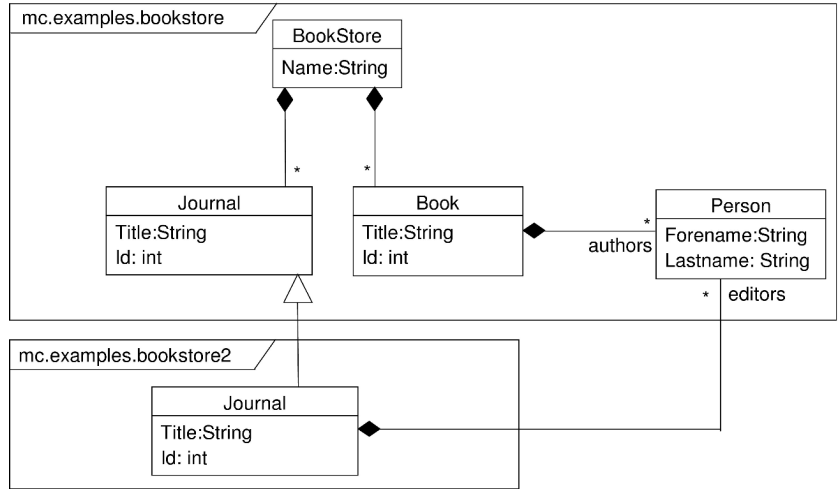

**Fig. 4.** Abstract syntax for the bookstore of Figure 3

This technique of overriding rules is typically used when the designer of the supergrammar does not foresee changes which are made in a subgrammar. However, there are scenarios where such modifications are predictable. In the case of our simple bookstore the developer may foresee that there will be subgrammars which introduce new items such as audio books. For these scenarios, MontiCore enables to define interfaces as possible extension point for subgrammars. Figure 5 shows a modified version of the basic grammar and the resulting abstract syntax.

Line 1 introduces a new interface `Item`, which is implemented by both `Book` and `Journal` (line 9 and 14 respectively). This definition leads to the generation of a Java-interface and the implements-relationship between the involved classes/interfaces. Note that we do not compute the attributes of interfaces automatically as the interfaces should serve as an extension point for new sublanguages which add new productions implementing this interface.

There are two main advantages of this version. First, a subgrammar can add new `Items` without changing or overriding the `Bookstore`. Therefore, a new production in the subgrammar simply implements the interface as described above. Second, the designer of the supergrammars can define attributes which have to be implemented by using the `ast` keyword that is used in order to modify the abstract syntax. Line 4 states that all implementing classes of the interface `Item` provide at least a title.

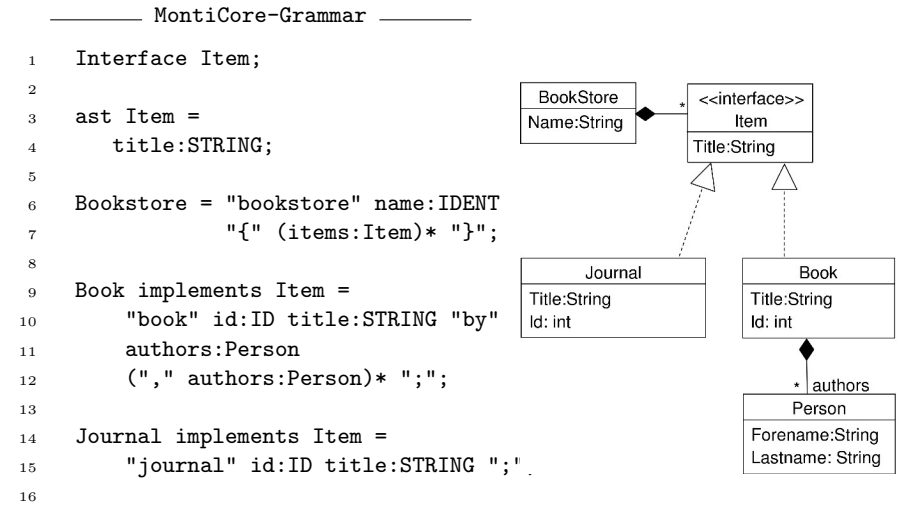

**Fig. 5.** Definition of bookstores using interfaces

Using this approach the designer of a sublanguage is able to add new kinds of items by extending our basic grammar and defining these new items as subtypes of the item interface as shown in Figure 6.

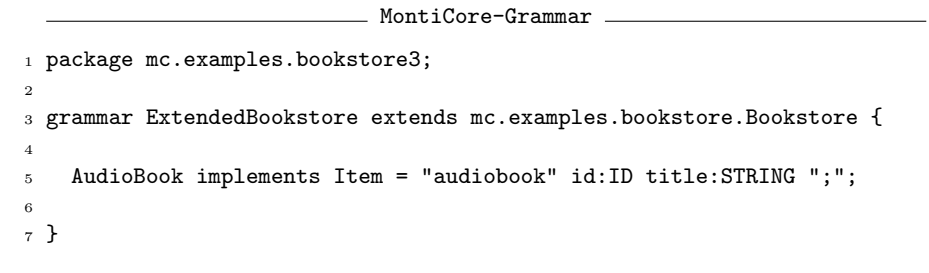

**Fig. 6.** Adding new items to the bookstore

In this simple example both grammars are not really decoupled from each other as the subgrammar directly inherits from the supergrammar. This might be a problem especially when only a few nonterminals of the subgrammar should be reused in other settings. In our example one can imagine that `AudioBooks` should be reused as items for a record shop. Using the former approach we had to change the subgrammar. Then, `AudioBooks` would implement another interface defined in the record shop grammar. However, this is often not desirable as the new grammar would be able to parse both book stores and record shops. In order to avoid this strong coupling MontiCore allows multiple grammar inheritance. This technique allows us to design both grammars separately by removing the inheritance between the grammars and the implements-relationship between `AudioBook` and `Item` and finally to define a third grammar which combines both supergrammars. Figure 7 shows an example. Please note that the grammar `mc.examples.audio.Audio` where the class `AudioBook` is defined is omitted here for space reasons because it is identical to the definition in Figure 6 except for the language and nonterminal inheritance.

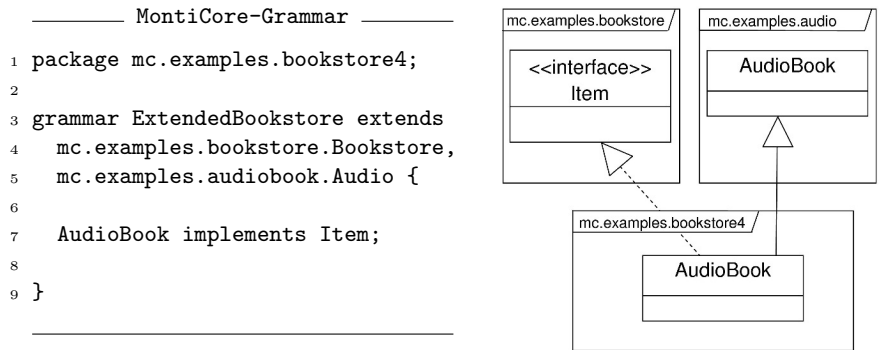

```
            ——— MontiCore-Grammar ———

1  package mc.examples.bookstore4;



3  grammar ExtendedBookstore extends
4    mc.examples.bookstore.Bookstore,
5    mc.examples.audiobook.Audio {



7    AudioBook implements Item;



9  }
```

**Fig. 7.** Multiple grammar inheritance

Language inheritance is typically used when the sublanguage is very similar to the superlanguage because otherwise problems in the lexical analysis may occur. The lexer rules of the subgrammar are a combination of the lexer rules of all supergrammars. They can be overridden in the same way like parser rules. The keywords of the language are a union of all keywords of the supergrammar plus the newly defined keywords of the subgrammar. A prominent example for the consequences can be seen in the introduction of the `assert` keyword in Java - legacy code using `assert` as identifier had to be adapted in order to conform to the new language. However, this problem can be avoided using language embedding where the recognition of the involved languages is strictly decoupled from each other because separate parsers and lexers are used.

### 4.2 Language Embedding

Domain specific languages are usually designed for a specific task; therefore it is often necessary to combine several languages to be able to define all properties of the desired software system precisely. A typical example for this approach is OCL which is used to define additional constraints on a model that cannot be expressed in the host language (e.g., class diagrams).

For convenience and for the sake of clarity it is desirable to write an OCL statement nearby the artifact it is constraining in the same file. Using standard approaches and parsing technologies would result in a monolithic and huge grammar combining both host language and OCL. This is even more problematic when combinations of more than two languages are used. Therefore, an independent development of all involved languages and a flexible combination mechanism is highly desirable.

MontiCore provides external nonterminals in grammars which means that their derivation is determined at configuration time by another appropriate language. We modify the bookstore example to be combined with an appropriate bibliography format (e.g., bibtex) as shown in Figure 8. In this example, line 1 introduces the external nonterminal **Bookentry** which is used on the RHS of **Book** (line 8). Note that there is no further information about the language to be embedded; hence the combination with a language that provides an arbitrary definition for BookEntry is valid.

---

**MontiCore-Grammar**

```
1    external Bookentry;
2    external Journalentry / example.IJournalEntry;
3
4    Bookstore = "bookstore" name:IDENT "{" (Book | Journal)* "}" ;
5
6    Book = "book" id:ID title:STRING
7           "by" authors:Person ("," authors:Person)*
8           Bookentry ";" ;
9
10   Journal = "journal" id:ID title:STRING Journalentry ";" ;
11
12   Person = forename:IDENT lastname:IDENT ;
```

---

**Fig. 8.** Expressing constraints for the embedded language by interfaces

The usage of external nonterminals leads to a composition relationship to **ASTNode** which is the base interface of all AST classes in MontiCore. However, it is sometimes desirable to define constraints for an embedded language in form of interfaces which must be implemented by the top level node of the embedded grammar. Therefore, MontiCore allows declaring the name of the interface to be implemented next to the definition of the external nonterminal as shown in line 2 of Figure 8. This version introduces an external nonterminal **Journalentry** that restrict the top level node of the embedded grammar to classes that implement **example.IJournalEntry**. The slash marks these interfaces as handwritten, therefore it is possible to access properties of the embedded language in form of methods from the host language. Furthermore, in this version the composition relationship is more reasonable since it is typed with **examples.IJournalEntry** instead of **ASTNode**. In addition, this example shows that the combination of languages is not restricted to two grammars.

In order to enable an independent development MontiCore derives parsers and AST classes separately and combines these components at configuration time as shown in Figure 9. For each grammar we generate a lexer and for each production of that grammar an adapter to the parser generated by Antlr without considering a concrete language combination. Therefore, each production can be used as a start production which is an important property to combine language fragments. This method enables us to reuse these artifacts without recompilation. Then, we combine the parsers/lexers-combinations to a superordinated parser which is able to switch between the different grammars. Every time a concrete parser finds an external nonterminal the control is passed to the superordinated parser which invokes the parser/lexer-combination of the embedded language.

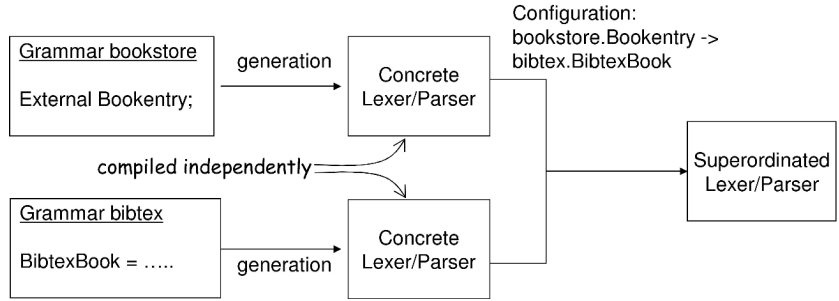

**Fig. 9.** Language embedding

Doing so, the bookstore parser ensures correct behavior by invoking the superordinated parser and thus, the bibtex parser (and lexer). Then, the rest of the text will be recognized according to that grammar. Thus, both languages are really independent from each other in their lexical and syntactic structure, and more important, the combination cannot be ambiguous as lexer and parsers are exchanged when switching the language. In order to allow compositional development beyond syntax, algorithms can be developed independently of each other as described in Section 5.

From a technical point of view a language combination consists of two pieces of information. The first one declares the grammar to be used when starting to recognize a text. The second piece of information in turn indicates which external nonterminal in a grammar should be replaced by which nonterminal of another grammar. Both can be expressed using a simple DSL as shown in Figure 10 or by combining the parsers/lexer by handwritten Java-code.

To realize the exchange of parsers and lexers in the desired fashion, we modified Antlr. First, all parsers/lexers use a shared queue which contains the input stream. Every time a parser consumes a token of length n, the first n characters are removed from the queue. It is important that this removal is not carried out when the lexer creates a token because it might be that the current lexer is not the right one for the currently considered part of the file. The main reason for this fact is that the position of the lexer is in front of the position of the parser in the file, especially when the parser uses its lookahead as this shifts the position of the lexer forward without shifting the current position of the syntactical analysis forward in the same ratio. Whenever the parser invokes the superordinated parser in order to recognize an embedded language, the queue is reset to the end

---
MontiCore-Grammar
---

```
1  // define rule Bookstore of the grammar mc.examples.bookstore.Bookstore
2  // to be used when starting to parse the text
3  mc.examples.bookstore.Bookstore.Bookstore bst <<start>>;
4
5  // embed bibtex rule for books as Bookentry
6  mc.examples.bibtex.Bibtex.BibtexBook bibBook in bst.Bookentry;
7
8  // embed bibtex rule for journals as Journalentry
9  mc.examples.bibtex.Bibtex.BibtexJournal bibJrn in bst.Journalentry;
```

---

**Fig. 10.** Combining languages using a special DSL

of the last token the host parser consumed. From this point onward, the embedded lexer starts to re-process the next characters. By this strategy we ensure the correct behavior of the embedded parser as no following token was typified by the host lexer. This strategy leads to repetitive lexing but does not impose great overhead as the maximal length of the re-typified character string is limited by the chosen lookhead or the considered length of syntactic or semantic predicates.

MontiCore grammars are a form of *grammar fragments* as defined in [Läm01]. Grammar fragments do not have a defined start production and referred nonterminals may be undefined. In the MontiCore context every grammar could be understood as a grammar fragment, because every production can be used as start production. This behavior is especially useful if parts of a language shall be reused as it was the case when we embedded a rule for `Bookentry` and another rule of the same grammar for `Journalentry`.

The example shown so far enables a language developer to embed different languages to different external nonterminals. Different sublanguages can be used but the decision is bound at configuration time and only a single language can be used for one external nonterminal. In Figure 11 we developed our running example further such that the developer can embed multiple languages for a single external nonterminal.

The different languages are registered under a unique name. Then, we assign string values to variables like shown in line 5 where we parse `booktype` and assign this value to the global variable `bt` (`astscript {set(bt,booktype);}`). In line 10 we refer to this variable in order to decide which language is used for `Bookentry`. In line 13 a different approach is taken in the sense that we call an embedded language depending on the value of a variable of the current rule instead of a global variable. Using this way a bookstore grammar can be developed that allows multiple formats for defining journals and books and the user of the resulting language can choose the one to use.

—————————— **MontiCore-Grammar** ——————————

```
1   external Bookentry;
2   external Journalentry / example.IJournalEntry;
3
4   Bookstore = "bookstore" name:IDENT
5                booktype:IDENT astscript { set(bt,booktype); }
6                "{" (Book | Journal)* "}" ;
7
8   Book = "book" id:ID title:STRING
9          "by" authors:Person ("," authors:Person)*
10         Bookentry<global bt> ";" ;
11
12  Journal = "journal" id:ID title:STRING
13             jt:IDENT Journalentry<jt> ";" ;
14
15  Person = forename:IDENT lastname:IDENT ;
```

**Fig. 11.** Using different parsers for a single nonterminal

## 5 Modular Development of Domain Specific Tools

The development of domain specific modeling languages consists not only of defining an abstract and a concrete syntax. Furthermore, there are several other steps necessary, e.g., code generation, syntactic checks, or the implementation of language specific tools. In the following we give two examples of how the further processing of modular languages is supported by MontiCore.

**Modular visitors.** Visitor-like algorithms simplify programming a code generator because each language construct can independently be translated as long as the tree structure of the AST is similar to the structure of the generated code. The main advantage of this design pattern is that the algorithm does not contain traversal code.

To increase the usability of the modular facilities within the MontiCore framework we complemented the modular language definition with a modular tree traversal. Without this facility a user would still program visitors for combinations of languages which are not reusable. Using our modular visitor concept, a user can program visitors independently for different fragments that handle the classes defined within a grammar. These classes can then be combined to a visitor without recompilation which invokes the different methods automatically. Thus, we use the same approach for combining visitors as for combining lexers and parsers in the case of language embedding.

For language inheritance the different visitors can be subtyped to change the behavior for the newly added or overwritten productions. Where subtyping is not feasible due to the single-supertype restriction Java imposes, delegation can be used.

**Modular tool generation.** Comfortable and usable tool support is an important success criterion for new domain specific modeling languages. However, the development of language specific tools such as editors is a time-consuming task, which gets even more complicated when we take the modularity concepts into account. Language embedding for example, requires not only an independent generation of parsers and lexers and a possibility to combine them at configuration time. The same approach should be reflected in tooling: independent development/generation of editors and afterwards, combination at configuration time. Doing so, it is possible to develop one editor for OCL separately and to combine it with other editors (e.g., for class diagrams, statecharts, or sequence diagrams) instead of implementing an editor for class diagrams with OCL, statecharts with OCL, and so on.

MontiCore grammars can be complemented by additional information which is used in order to generate language specific editors as Eclipse plugins. Generated editors offer different comfort functionalities such as syntax highlighting, an outline, foldable code regions, or error messages. We will only briefly introduce editor generation in this paper. For deeper discussion on editor generation we refer to [KRV07a].

The editor generation of MontiCore is fully integrated with the modularity concepts we discussed in this paper. For each grammar (which possibly has external nonterminals) we generate visitors as described in the beginning of this section. These visitors evaluate those parts of the abstract syntax which are defined in the current grammar. Among other things, they are able to provide an outline page with items, or to advise the document which code regions can be folded. Then, we are able to combine the visitors without recompilation in order to gain an editor which supports the current language combination (e.g., class diagrams and OCL). Furthermore, when handling grammar inheritance, we simply use object oriented inheritance between the generated editors. This allows us to override (or to complement) the behavior of the supereditor, and even more important, to further develop the supereditor without a need to recompile every subeditor, and thus, every sublanguage.

## 6 Conclusion and Outlook

In this paper we explained how the DSL framework MontiCore and especially how the defining grammar format has been extended to support two different kinds of modularity mechanisms. The modular development as well as the combination and extension techniques simplify the integration of DSLs and their according tools in software development projects.

*Language inheritance* can be used to extend existing languages by new nonterminals or to override existing nonterminals of the supergrammar. The effects on the abstract syntax are similar to those of UML package merge as only the delta is used to generate new abstract syntax classes. In the case of overriding, these new classes directly inherit from the old ones. Therefore, existing artifacts such as symbol tables, context constraint checkers, or code generation can be reused with minor modifications. *Language embedding* on the other hand is useful to explicitly add points of variation in modeling languages to include fragments of other languages. The strong decoupling of the languages by separate lexers/parsers minimizes interferences and permits a component based composition of grammars.

In addition, MontiCore supports the development of modular domain specific modeling languages beyond syntax. For this purpose, we explained how a language definition is complemented with further concepts like editor generation and tree traversal. Both concepts reflect the modular approaches for defining the abstract and the concrete syntax of a language.

MontiCore has been used to define several small DSLs as well as some complex languages. We realized a subset of UML [Rum04b, Rum04a] including statecharts, class, sequence, and object diagrams. Furthermore, we developed a grammar for OCL and Java 5 and helped industrial partners to implement a DSL for a part of an AUTOSAR specification [Höw07]. In addition, we evaluated view based modeling of logical automotive architectures [GHK+08] with MontiCore. In addition, MontiCore itself is realized using a bootstrapping approach. Currently about 75% of the code is generated from several DSLs.

We currently elaborate on a possible connection of the MontiCore framework with EMF [BSM+03] as well as MOFLON [AKRS06] to simplify the interoperability of DSLs defined with MontiCore and metamodeling techniques. This will allow us to gain further support for model transformations and code generation from a variety of tools. Furthermore, we will explore composition of languages beyond syntax. This includes compositional context constraints and symbol tables as this plays an important role especially in the case of language embedding.

*Acknowledgment:* The work presented in this paper is undertaken as a part of the MODELPLEX project. MODELPLEX is a project co-funded by the European Commission under the "Information Society Technologies" Sixth Framework Programme (2002-2006). Information included in this document reflects only the authors' views. The European Community is not liable for any use that may be made of the information contained herein.